\documentclass[aps,twocolumn,preprintnumbers,nofootinbib,superscriptaddress]{revtex4-2}

\pdfoutput=1

\usepackage[usenames,dvipsnames,table]{xcolor}
\usepackage{graphicx,amsmath,amssymb,amsthm,multirow,array,bm,soul}
\usepackage[mathscr]{eucal}
\usepackage[bbgreekl]{mathbbol}
\usepackage{amsfonts}
\usepackage{hyperref}
\usepackage{tikz}
\usepackage{float}
\usetikzlibrary{3d,calc}

\newcommand{\beq}{\begin{equation}}
\newcommand{\eeq}{\end{equation}}

\def\<{\langle}
\def\>{\rangle}

\newcommand{\llangle}{{\langle\!\langle}}
\newcommand{\rrangle}{{\rangle\!\rangle}}

\begin{document}

\title{Defect Localized Entropy: Renormalization Group and Holography}

\author{Ma-Ke Yuan}
\email{mkyuan19@fudan.edu.cn}
\affiliation{Department of Physics and Center for Field Theory and Particle Physics, Fudan University, Shanghai 200433, China}
\author{Yang Zhou}
\email{yang\_zhou@fudan.edu.cn}
\affiliation{Department of Physics and Center for Field Theory and Particle Physics, Fudan University, Shanghai 200433, China}
\affiliation{State Key Laboratory of Surface Physics, Fudan University, Shanghai 200433, China}
\affiliation{Peng Huanwu Center for Fundamental Theory, Hefei, Anhui 230026, China}

%\author{}
%\email{}
%\affiliation{}

%\date{\today}

\begin{abstract}
We consider $p$-dimensional defects in $D$-dimensional conformal field theories (CFTs) and construct defect localized entropy by performing Casini-Huerta-Myers transformation for the system with defect. The defect localized entropy is a measure of entanglement between the degrees of freedom localized on the defect. We show that at the fixed point of defect renormalization group (RG) flow, defect localized entropy is equal to minus defect free energy for universal part. We construct defect $C$-functions from the defect localized entropy for surface defects and volume defects, and show that they monotonically decrease in both cases following the entropic method by Casini and Huerta. We also study the holographic dual of defect localized entropy and find that it is given by the minimal surface located at string or brane worldvolume embedded in the holographic bulk.
\end{abstract}

\maketitle

%%%%%%%%%%%%%%%%%%%%%%%%%%%%%%%%%%%%%%%%%%%%%%%%%%%%%%%%%%%%%%%%%%%%%%%%%%%%%%%%%%%%%%%%%%%%%%%
\subsection{Introduction}

Defects are usually defined by non-local operators with fixed spacetime location in QFT. Therefore they can be classified by their dimensions, such as line defects, surface defects, etc. The familiar examples of defects are Wilson lines and Wilson surfaces in gauge theory. Not all defects can be described by operators in terms of bulk elementary fields. A large class of defects are defined through boundary conditions. For instance, boundary or interface can be viewed as codimension one defects. Local operators in QFT can be regarded as zero-dimensional defects although we usually do not treat them in this way.

Counting the degrees of freedom under RG flow is of great importance in QFT. Zamolodchikov proved the existence of a $C$-function which monotonically decreases under RG flows and coincides with the CFT central charge at the conformal fixed point in $D = 2$~\cite{Zamolodchikov:1986gt}. Further results in diverse dimensions were discussed in~\cite{Cardy:1988cwa, Cappelli:1990yc, Osborn:1991gm, Casini:2004bw, Casini:2006es, Myers:2010xs, Myers:2010tj, Jafferis:2010un, Jafferis:2011zi, Klebanov:2011gs, Komargodski:2011vj, Myers:2012ed, Liu:2012eea, Casini:2012ei, Elvang:2012st, Elvang:2012yc, Yonekura:2012kb, Antipin:2013pya, Grinstein:2013cka, Jack:2013sha, Baume:2014rla, Grinstein:2014xba, Giombi:2014xxa, Kawano:2014moa, Cordova:2015vwa, Jack:2015tka, Cordova:2015fha, Casini:2015woa, Pufu:2016zxm, Casini:2017vbe, Lashkari:2017rcl, Fluder:2020pym, Delacretaz:2021ufg}. For our purposes we highlight the entropic method initiated by Casini and Huerta in~\cite{Casini:2006es, Casini:2012ei}, which establishes the $C$-theorem using the entanglement entropy (EE) across a spherical entangling surface which divides the space into two parts on a time slice. 

Let us now consider QFT with defects. A defect RG flow may be triggered by perturbing the defect CFT (DCFT) with relevant defect operators. One natural question is if there exists a defect $C$-function, which counts the defect degrees of freedom. Defect RG have been studied in~\cite{Affleck:1991tk, Ludwig:1994nf, Dorey:1999cj, Yamaguchi:2002pa, Friedan:2003yc, Azeyanagi:2007qj, Takayanagi:2011zk, Fujita:2011fp, Nozaki:2012qd, Estes:2014hka, Gaiotto:2014gha, Jensen:2015swa, Casini:2016fgb, Kobayashi:2018lil, Casini:2018nym, Giombi:2020rmc, Wang:2020xkc, Nishioka:2021uef, Wang:2021mdq, Sato:2021eqo, Cuomo:2021rkm, Cuomo:2021kfm, Cuomo:2022xgw}. For our purposes, it is important to highlight the conjecture by Kobayashi, Nishioka, Sato and Watanabe~\cite{Kobayashi:2018lil} that a defect $C$-function should coincide (up to a sign) with defect free energy at fixed point. In the case of line defects, Cuomo, Komargodski and Raviv-Moshe~\cite{Cuomo:2021rkm} proved the monotonicity of an entropy formula.

The focus of this paper is the construction of defect $C$-functions for various dimensional defects. We first define defect localized entropy by counting entanglement between degrees of freedom localized on the defect. Here by defect degrees of freedom we did not mean the operator localized at the defect, rather we refer to degrees of freedom in the internal Hilbert space. For instance, Wilson loop can be equivalently represented as 1$d$ fermion or boson path integral. We call those fermion or boson as defect degrees of freedom. By construction, Wilson loop operator is recovered by integrating out the fermion/boson~\cite{Gomis:2006sb, Hoyos:2018jky}. In this paper we are interested in computing entanglement in the internal Hilbert space. When we quantize the internal degrees of freedom, the bulk fields are treated as classical background fields or potentials~\cite{tong2018gauge}. The bulk path integral provides a distribution for the classical background fields. Therefore the system looks like an ensemble. Essentially we want to compute the entanglement entropy for the defect state after integrating out the bulk. 

Notice that the entanglement entropy we defined here is different from that defined in~\cite{Jensen:2013ora, Lewkowycz:2013laa}. The latter is given by the defect contribution to the bulk EE and generally not a decreasing function along defect RG flow as shown in~\cite{Kobayashi:2018lil}. Employing Casini-Huerta-Myers map (CHM map), we show that defect localized entropy equals to minus defect free energy for universal terms at fixed points of defect RG. We will construct defect $C$-functions from the defect localized entropy for surface defects and volume defects, and show that they monotonically decrease in both cases following the quantum information approach initiated by Casini and Huerta. We also discuss the holographic dual of defect localized entropy.

%%%%%%%%%%%%%%%%%%%%%%%%%%%%%%%%%%%%%%%%%%%%%%%%%%%%%%%%%%%%%%%%%%%%%%%%%%%%%%%%%%%%%%%%%%%%%%%
%%%%%%%%%%%%%%%%%%%%%%%%%%%%%%%%%%%%%%%%%%%%%%%%%%%%%%%%%%%%%%%%%%%%%%%%%%%%%%%%%%%%%%%%%%%%%%%

\subsection{Setup of DCFT}
We consider a local, unitary, Euclidean CFT on a $D$-dimensional spacetime $\cal{M}$ with coordinates $x^{\mu}$ ($\mu = 0, \dots, D - 1$) and metric $g_{\mu\nu}$, the so called ``bulk'' CFT. We introduce a codimension $D - p$ defect along a $p$-dimensional submanifold $\Sigma$ with coordinates $\hat x^a$ ($a = 0, \dots, p - 1$) and induced metric $\gamma_{ab} \equiv g_{\mu\nu} \partial_a X^{\mu} \partial_b X^{\nu}$, where $X^{\mu}(\hat x)$ is the embedding function parameterizing $\Sigma \hookrightarrow \cal{M}$. Physically, the defect can arise from coupling $p$-dimensional degrees of freedom to the bulk CFT.\footnote{Or from imposing boundary conditions on the bulk CFT fields. In the codimension one case, if one of the two sides is trivial, the defect becomes a boundary.} As mentioned in the introduction, we treat $p$-dimensional degrees of freedom as internal degrees of freedom, which means that one should integrate out them to obtain the defect operator. If a Lagrangian description exists, the DCFT action has the ambient part and the defect part
\begin{equation}\label{eq:action}
I_{\text{DCFT}} = \int \text{d}^D x \sqrt{g} {\cal L}_{\text{CFT}}[\phi] + \int \text{d}^p \hat x \sqrt{\gamma} {\cal L}_{\text{defect}}[\phi, \psi]\ , 
\end{equation}
where $\phi$ denotes the bulk degrees of freedom and $\psi$ the defect degrees of freedom.\footnote{See \cite{tong2018gauge} for such an explicit construction of Wilson loop operator in gauge theory.}
%%%%%%%%%%%%%%%%%%%%%%%%%%%%%%%%%%%%%%%%%%%%%%%%%%%%%%%%%%%%%%%%%%%%%%%%%%%%%%%%%%%%%%%%%%%%%%%
%%%%%%%%%%%%%%%%%%%%%%%%%%%%%%%%%%%%%%%%%%%%%%%%%%%%%%%%%%%%%%%%%%%%%%%%%%%%%%%%%%%%%%%%%%%%%%%

\textit{Stress tensor.}~Let us restrict our attention to conformal defects, which are hyperplanes or spheres, to preserve part of the conformal symmetry. A $p$-dimensional conformal defect breaks the ambient conformal symmetry $SO(1, D + 1)$ to $SO(1, p + 1) \times SO(D - p)$. For a CFT in flat space, conformal symmetry forces $\langle T_{\mu\nu}(x) \rangle = 0$. However, in the presence of defect, the one-point function of ambient stress-energy tensor does not necessarily vanish. To illustrate, consider a $p$-dimensional planar defect in $\mathbb{R}^D$. The metric is then divided into parallel and transverse directions: $\text{d}s^2 = \text{d} \hat x^a \text{d} \hat x^a + \text{d}x^i \text{d}x^i$ with $a = 0, \dots, p - 1$ and $i = p, \dots, D - 1$. The stress-energy tensor follows from varying the defect partition function and it is often useful to split it into the ambient part $T^{\mu\nu}$ and the defect localized part $t^{ab}$. See for instance~\cite{Billo:2016cpy, Kobayashi:2018lil}. The ambient stress-energy tensor is a symmetric traceless tensor of dimension $D$ and spin $2$, hence the (partial) conservation plus residual conformal symmetry fix its form completely~\cite{Kapustin:2005py, Billo:2016cpy}\footnote{In this paper the notation $\llangle O \rrangle$ refers to the correlation function measured in the presence of defect
\begin{equation*}
\llangle O \rrangle \equiv \langle O D \rangle / \langle D \rangle\ ,
\end{equation*}
where $D$ denots the defect. }
\begin{equation}\label{Tbulk}
\begin{split}
\llangle T^{ab} \rrangle &= -\frac{D - p - 1}{D} \frac{h}{|x^i|^D} \delta^{ab}\ , \quad \llangle T^{ai} \rrangle = 0\ ,\\
\llangle T^{ij} \rrangle &= \frac{h}{|x^i|^D} \left(\frac{p + 1}{D}\delta^{ij} - \frac{x^i x^j}{|x^i|^2}\right)\ ,
\end{split}
\end{equation}
where $h$ characterizes the property of the defect. Furthermore, at the fixed point of defect RG flow, the defect localized stress tensor $t^{ab}$ is a defect local operator of dimension $p$ whose vev must vanish due to the residual conformal symmetry on the defect~\cite{Kobayashi:2018lil}
\begin{equation}\label{TD}
\llangle t^{ab} \rrangle = 0\ .
\end{equation}
This does not hold if we are away from the fixed point of defect RG flow.

%%%%%%%%%%%%%%%%%%%%%%%%%%%%%%%%%%%%%%%%%%%%%%%%%%%%%%%%%%%%%%%%%%%%%%%%%%%%%%%%%%%%%%%%%%%%%%%
%%%%%%%%%%%%%%%%%%%%%%%%%%%%%%%%%%%%%%%%%%%%%%%%%%%%%%%%%%%%%%%%%%%%%%%%%%%%%%%%%%%%%%%%%%%%%%%
\subsection{Defect localized entropy}
Now we define defect localized entropy based on the previous setup. Consider a $p-2$ dimensional sphere of radius $\ell$ localized on the static defect which divides the defect into two parts. We want to compute the EE between the two parts for the defect state constructed by integrating out the bulk, which is von Neumann entropy 
\begin{equation*}
S = -\text{Tr}(\rho\log\rho)
\end{equation*} of the defect reduced density matrix $\hat{\rho}_A$ constructed by only cutting along the defect subregion $A$
\begin{equation}\label{densityM}
\begin{split}
[\hat{\rho}_A]_{ab} = &\frac{1}{Z^\text{DCFT}} \int_{\cal{M}} \mathcal{D}\phi \int_{\Sigma} \mathcal{D}\psi\ e^{- I_{\text{CFT}} - I_{\text{defect}}}\\ 
&\times \prod_{\hat{x} \in A} \delta (\psi(0^+, \hat{x}) - \psi_b(\hat{x})) \delta (\psi(0^-, \hat{x}) - \psi_a(\hat{x}))\ , 
\end{split}
\end{equation}
where $\phi$ denotes the bulk degrees of freedom and $\psi$ the defect degrees of freedom (more details about this density matrix are given in appendix~\ref{appx-PI}). The defect localized entropy can be considered as a correlation measure for the defect degrees of freedom $\psi$. And it can also be viewed as a generalization of the ordinary EE to the defect. As we will see, this generalization defines an intrinsic property of the defect and quantifies the defect degrees of freedom. 

\begin{figure}[htbp]
\centering
\includegraphics[height=3.7cm]{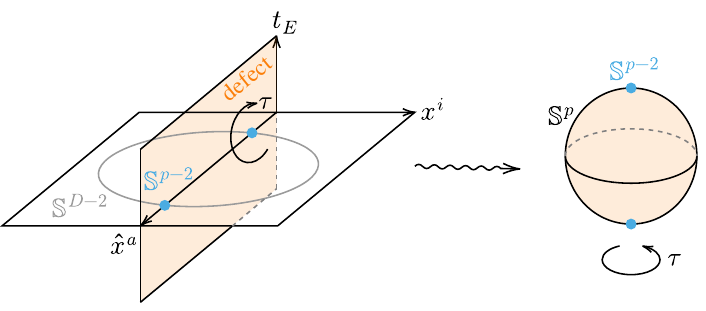}
\caption{Illustration of the Casini-Huerta-Myers map \eqref{coordT}. }
\label{fig-Trans}
\end{figure}

As usual, the entanglement entropy can be computed by replica trick. Let us start by considering a codimension two bulk sphere ($D-2$ dimensional) of the same radius $\ell$ centered at the defect, with the previous $p-2$ sphere a subsphere, as shown in figure~\ref{fig-Trans}. Following~\cite{Casini:2011kv, Jensen:2013lxa, Kobayashi:2018lil, Jensen:2018rxu} we perform CHM map for our system. We parameterize the flat metric as
\begin{equation}\label{flatmetric}
\text{d}s^2 = \text{d}t_E^2 + \text{d}\hat x^a d\hat x^a + (\text{d}|x^i|)^2 + |x^i|^2 \, \text{d}s_{\mathbb{S}^{D-p-1}}^2,
\end{equation}
with the defect along $\{t_E, \hat x^a\}$ (here $a = 1, \dots, p-1$) and located at $|x^i| = 0$. Defining $r^2 = |\hat x^a|^2 + |x^i|^2$, the coordinate transformation
\begin{equation}\label{coordT}
\begin{split}
	t_E = \frac{\ell\cos\theta \sin\left(\frac{\tau}{\ell}\right)}{1+\cos\theta\cos\left( \frac{\tau}{\ell}\right)}\ ,& \quad r = \frac{\ell \sin\theta}{1 + \cos\theta \cos\left(\frac{\tau}{\ell}\right)},\\
	|\hat x^a| = r \cos\phi\ ,& \quad |x^i| = r \sin\phi\ ,
\end{split}
\end{equation}
maps the $D$ dimensional flat space to a $D$ dimensional sphere $\mathbb{S}^D$,
\begin{align}
\label{E:SD}
\widetilde \Omega^2 \text{d}s^2 &= \cos^2\theta \text{d}\tau^2 + \ell^2(\text{d}\theta^2 + \sin^2\theta \text{d}\Omega^2)\ ,\\
\text{d}\Omega^2 &= \left( \cos^2\phi \text{d}s^2_{\mathbb{S}^{p-2}} + \text{d}\phi^2 + \sin^2\phi \text{d}s^2_{\mathbb{S}^{D-p-1}} \right)\ ,
\end{align}
where $\widetilde \Omega  = 1 + \cos\theta \cos(\tau/\ell)$, $\tau \in (0, 2\pi]$, $\theta \in [0, \pi/2]$, and $\phi \in [0, \pi/2]$. %The defect is then located at $\phi = 0$, i.e. along a maximal dS$_p$.
%maps the bulk sphere's causal development to the static patch of $D$-dimensional de Sitter space, dS$_{D}$, with metric
%\begin{align}
%\label{E:dS}
%\widetilde \Omega^2 \text{d}s^2 &= -\cos^2\theta \text{d}\tau^2 + \ell^2(\text{d}\theta^2 + \sin^2\theta \text{d}\Omega^2)\ ,\\
%\text{d}\Omega^2 &= \left( \cos^2\phi \text{d}s^2_{\mathbb{S}^{p-2}} + \text{d}\phi^2 + \sin^2\phi \text{d}s^2_{\mathbb{S}^{D-p-1}} \right)\ ,
%\end{align}
%where $\widetilde \Omega  = 1 + \cos\theta \cosh(\tau/\ell)$, $\tau \in (-\infty, \infty)$, $\theta \in [0, \pi/2]$, and $\phi \in [0, \pi/2]$. The defect is then located %at $\phi = 0$, i.e. along a maximal dS$_p$.
%Analytically continuing to Euclidean time, $\tau \rightarrow - i\tau_E$ with $\tau_E \sim \tau_E + 2\pi\ell$, \eqref{E:dS} becomes the metric of $\mathbb{S}^D$, with
Now the defect is located at $\phi = 0$, wrapping a maximal $\mathbb{S}^p$ as illustrated in figure~\ref{fig-Trans}. The defect free energy can be computed
\begin{equation}
\label{eq:dsfree}
F_D = - \log \left( Z^\text{DCFT}/Z^\text{CFT} \right) \equiv -\log\langle D[\mathbb{S}^p] \rangle\ ,
\end{equation}
where $Z^\text{DCFT}$ is the DCFT's Euclidean partition function on $\mathbb{S}^D$ and $D[\mathbb{S}^p]$ denotes the defect operator on the $p$-dimensional subsphere.\footnote{Whenever we talk about the defect operator, the defect(internal) degrees of freedom have been integrated out.} Under the replica trick, the $p-2$ dimensional entangling surface, which becomes $p-2$ dimensional sphere now, should be the fixed point. 
This means that we only replica the defect operator while keeping the bulk intact. The procedure can be understood as a construction of a new defect operator by inserting a $p-2$ dimensional twist operator at the entangling surface within the defect. Recalling the ordinary definition of EE by the continuation $n\rightarrow 1$ in the replica trick, one can therefore define the defect localized EE by taking $n\rightarrow 1$,
\begin{equation}\label{SD0}
S_{D} = \lim_{n \rightarrow 1} \frac{1}{1 - n} \log\frac{\langle D[\mathbb{S}_n^p] \rangle}{\langle D[\mathbb{S}^p] \rangle^n}\ .
\end{equation}
Obviously the definition \eqref{SD0} expressed in terms of $D$ operators, is a formula obtained from bulk point of view. As mentioned before, the defect localized EE as von Neumann entropy for the defect degrees of freedom, can also be defined at the level of the reduced density matrix~(\ref{densityM}). In appendix~\ref{appx-PI} we demonstrate that the two definitions agree. To compute \eqref{SD0}, let us consider the $n - 1$ deformation as the deformation of the $\tau\tau$ component of the metric, and expand the defect free energy,
\begin{equation}
\log \langle D[\mathbb{S}_n^p] \rangle = \log \langle D[\mathbb{S}^p] \rangle - \frac{1}{2} \int_{\mathbb{S}^p} \delta \gamma_{\tau\tau} \llangle t^{\tau\tau} \rrangle + \cdots\ , 
\end{equation} 
where $\delta \gamma^\tau{}_\tau = \gamma^{\tau\tau} \delta \gamma_{\tau\tau} = n^2 - 1$. Since the higher orders do not contribute to the entropy we obtain 
\begin{equation}\label{SD}
S_{D} = \log \langle D[\mathbb{S}^p] \rangle\ + \int_{\mathbb{S}^p} \llangle t^\tau{}_\tau \rrangle\ .
\end{equation}
The first term and the second term are the expectation value of the defect and the Killing energy (times $\beta = 2 \pi \ell$) localized on the defect.\footnote{Note that the possible anomalous contribution from the background when the bulk dimension is even has already been deducted.} This is our main result. An alternative derivation of \eqref{SD} based on the relation between EE in flat space and thermal entropy on sphere under CHM map~\cite{Casini:2011kv} is given in appendix~\ref{appx-CHM}. 

We thus need $\llangle t^\tau{}_\tau \rrangle$, which can be obtained by Weyl transformation from the flat-space $\llangle t^{ab} \rrangle$ in \eqref{TD}. When $p$ is odd, there is no Weyl anomaly therefore $\int_{\mathbb{S}^p} \llangle t^\tau{}_\tau \rrangle = 0$. When $p$ is even, there is a contribution from $t^{ab}$'s anomalous Weyl transformation law. In this case, the second term in \eqref{SD} is finite, but the first term $\log \langle D[\mathbb{S}^p] \rangle$ diverges. Focusing on the universal part, we find a relation between defect localized EE and defect free energy at defect fixed point,
\begin{equation}\label{SFD}
S_{D} = \log \langle D[\mathbb{S}^p] \rangle = -F_D\ ,
\end{equation}
for both even and odd dimensional defects. This is really the analogy of the universal relation between EE and sphere free energy for bulk CFTs worked out by Casini, Huerta and Myers~\cite{Casini:2011kv}. We emphasize that the relation \eqref{SFD} does not hold if we leave from the fixed point of defect RG flow.

%%%%%%%%%%%%%%%%%%%%%%%%%%%%%%%%%%%%%%%%%%%%%%%%%%%%%%%%%%%%%%%%%%%%%%%%%%%%%%%%%%%%%%%%%%%%%%%
%%%%%%%%%%%%%%%%%%%%%%%%%%%%%%%%%%%%%%%%%%%%%%%%%%%%%%%%%%%%%%%%%%%%%%%%%%%%%%%%%%%%%%%%%%%%%%%
\subsection{\boldmath A defect $C$-function\unboldmath }
\textit{Line defect}~Following the previous setup, the line defect ($p=1$) is along $t_E$ and located at $r= 0$ in the flat space (\ref{flatmetric}). 
After the CHM map \eqref{coordT}, the defect is now along a maximal $\tau$ circle with radius $\ell$ in $\mathbb{S}^{D}$. Since there is no Weyl anomaly for line defect, the defect localized energy vanishes at the defect fixed point, and the defect localized EE is given by minus the defect free energy 
\begin{equation}
\label{eq:DLEE1}
S_D^{\rm p=1} = -F_D = \log \left(Z^\text{DCFT}/Z^\textrm{CFT}\right) = \log\langle L[\mathbb{S}^1]\rangle \ .
\end{equation}
Along the defect RG flow, $F_D$ depends on $\beta=2\pi\ell$ through the relevant deformation along the defect. Under replica trick, it is easy to show that $\beta\propto n$, and the defect localized EE \eqref{SD0} is given by
\begin{equation}
\begin{split}
S_D^{\rm p=1} 
&= (1-n\partial_n)\log\langle L[\mathbb{S}^1_n]\rangle\big|_{n\to 1}\\
&=(1-\beta\partial_\beta)\log\langle L[\mathbb{S}^1]\rangle\ .
\end{split}
\end{equation}
This coincides with the line defect entropy formula, proven to be a monotonic defect $C$-function in general $D$-dimensional bulk CFT~\cite{Cuomo:2021rkm}. 

\textit{Surface defect}~For surface defects ($p=2$) at the fixed point, the defect free energy \eqref{eq:dsfree} can be computed, $Z^\text{DCFT}/Z^\text{CFT} \propto (\ell / \epsilon)^{b / 3}$, where $b$ is the defect central charge~\cite{Jensen:2015swa}. The defect localized EE is therefore given by\footnote{The Weyl anomaly term is given by eq.(3.10) in \cite{Kobayashi:2018lil}, $\int_{\mathbb{S}^2} \llangle t^\tau{}_\tau \rrangle = -b/3$, which does not rely on $\ell$ thus does not modify the refined formula \eqref{refine2}. Therefore, we put it into the term ${\cal O}(\varepsilon^0)$ of \eqref{eq:DLEE2}. }
\begin{equation}
\label{eq:DLEE2}
S_D^{\rm p = 2} = \frac{b}{3}\log \left( \frac{\ell}{\varepsilon} \right) + {\cal O}(\varepsilon^0)\ .
\end{equation}
A refined formula will pick up the universal coefficient,
\begin{equation}\label{refine2}
3\ell\partial_\ell S_D^{\rm p=2} = b\ .
\end{equation}
When there is a relevant deformation along the defect, there will be corrections for \eqref{eq:DLEE2} depending on $\ell$ and the refinement $3\ell\partial_\ell$ does not give the central charge any more. 

Motivated by~\cite{Casini:2006es}, for a generic defect RG flow, it is tempting to conjecture that {\it $3\ell\partial_\ell S_\mathcal{D}^{\rm p=2}$ as a defect $C$-function, is monotonically decreasing along the defect RG flows.} Here we use $S_\mathcal{D}$ to denote the defect localized EE defined on planar defect in flat space, to distinguish from $S_D$, which is defined on spherical defect after the CHM map~\eqref{coordT}. Their universal parts are equal at the fixed point as discussed above, but they differ at a generic point along the defect RG flow.\footnote{This may explain the discrepancy between the perturbative monotonicity claimed in the recent paper~\cite{Shachar:2022fqk} and the non-perturbative monotonicity of $S_\mathcal{D}$. After this paper was submitted to arXiv, Shachar, Sinha and Smolkin propose \textit{the renormalized defect entropy} in~\cite{Shachar:2022fqk} and show its monotonicity near the fixed point following the field theoretical approach in~\cite{Cuomo:2021rkm}. The renormalized defect entropy is given by eq.(3.4) in~\cite{Shachar:2022fqk}
\begin{equation*}
S = R\partial_R \left( 1 - \frac{1}{2}R\partial_R \right) \log \frac{Z^\text{DCFT}}{Z^\text{CFT}}\ , 
\end{equation*}
where the term $\left( 1 - \frac{1}{2}R\partial_R \right) \log \frac{Z^\text{DCFT}}{Z^\text{CFT}}$ is in fact the same as defect localized entropy $S_D$ we define in this paper.} Notice that the refinement in \eqref{refine2} is the same as that employed in $2D$ QFT to define an entropic $C$-function~\cite{Casini:2006es}.
%The $R\partial_R$ picks up the trace of stress tensor, but in our definition \eqref{SD} we only need the $\tau\tau$ component, thus there is a factor $1/2$. } 

\begin{figure}[htbp]
\centering
\includegraphics[height=2.7cm]{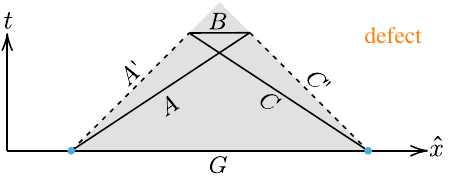}
\caption{A proof of the $p = 2$ defect $C$-theorem \textbf{(working on defect)}. This triangle is the causal development of $G$ and is placed on the \textbf{defect}. The proper lengths are $L(G) = \ell$, $L(B) = r$, and $L(A) = L(C) = \sqrt{r\ell}$. }
\label{fig-c-theorem}
\end{figure}

Now we give an information theoretical proof of the above conjecture following Casini and Huerta. Let us go back to the Minkowski space with the planar defect. Following the quantum information approach in~\cite{Casini:2006es}, we consider the quantum state associated with the defect degrees of freedom~(\ref{densityM}) and employ Lorentz invariance, unitarity, causality and the strong subadditivity of defect localized EE. Instead of tuning the energy scale $\mu$, here we trigger the defect RG flow by tuning the system size $\ell$. Thus the defect $C$-theorem is equivalently stated as follow: $c(\ell)$ is a monotonically decreasing function under RG flows.
As illustrated in figure~\ref{fig-c-theorem}, intervals $A'$ and $C'$ are on the light cone (the dashed lines) and intervals $B$ and $G$ lie on the time slice with their lengths $r$ and $\ell$ respectively. We emphasize that all the boosted intervals (or disks in appendix~\ref{appx-ctheorem}, and their intersections/unions) are located on the defect because we study the defect state after integrating out the bulk. The strong subadditivity of defect localized EE gives
\begin{equation}
S_\mathcal{D}^{A' \cup B \cup C'} + S_\mathcal{D}^{B} \leq S_\mathcal{D}^{A' \cup B} + S_\mathcal{D}^{B \cup C'}\ .   
\end{equation}
Using unitarity, we are able to identify the defect localized EE of $A' \cup B$ and that of $A$. Using Lorentz invariance, we can further write the defect localized EE of $A$ as a function of its proper length. Thus we have
\begin{equation}\label{eq-subadditivity}
\begin{split}
&S_\mathcal{D}^{G} + S_\mathcal{D}^{B} \leq S_\mathcal{D}^{A} + S_\mathcal{D}^{C}\\ 
\Rightarrow\ &S_\mathcal{D}(\ell) + S_\mathcal{D}(r) \leq S_\mathcal{D}(\sqrt{r\ell}) + S_\mathcal{D}(\sqrt{r\ell})\ .  
\end{split} 
\end{equation}
Performing Taylor expansion of \eqref{eq-subadditivity} at $\ell$ with respect to $\epsilon = r - \ell$ to the second order, one gets 
\begin{equation}\label{eq-taylor}
\begin{split}
&S_\mathcal{D}(\ell) + S_\mathcal{D}(\ell) + \epsilon S'_\mathcal{D}(\ell) + \frac{\epsilon^2}{2} S''_\mathcal{D}(\ell)\\ 
\leq\ &2S_\mathcal{D}(\ell) + 2\left( \frac{\epsilon}{2} - \frac{\epsilon^2}{8\ell} \right)S'_\mathcal{D}(\ell) + \frac{\epsilon^2}{4} S''_\mathcal{D}(\ell)\ , 
\end{split}
\end{equation}
which is actually what we want
\begin{equation}
\left( 3 \ell \partial_\ell S_\mathcal{D}(\ell) \right)' = 3 S'_\mathcal{D}(\ell) + 3 \ell S''_\mathcal{D}(\ell) \leq 0\ . 
\end{equation}
A $b$-theorem, stated that $b_\text{UV}\geq b_\text{IR}$, has been proven by Jensen and O'Bannon previously~\cite{Jensen:2015swa}. Our result is consistent with their result and provides a $C$-function for the entire defect RG flow.

Let us discuss the physical conditions we impose for the defect state.\footnote{These conditions are also imposed for $p=3$.} While Lorentz invariance and unitarity are quite reasonable assumptions, we do assume that the defect states also evolve causally.\footnote{It would be interesting to sharpen the causality condition and use it to classify defects.} Further more, strong subadditivity is true for any entanglement entropy.

\textit{Volume defect}~For volume defects ($p=3$), there is no Weyl anomaly and at the defect fixed point the localized energy vanishes. We therefore obtain the defect localized EE given by \eqref{SD} for $p=3$,
\begin{equation}
\label{eq:DLEE3}
S_D^{\rm p=3} = \log\langle D[\mathbb{S}^3]\rangle\ .
\end{equation}
Like sphere free energy of CFT$_3$, (\ref{eq:DLEE3}) may carry a linear divergence. Away from the defect fixed point, a refined formula is required to define a proper finite $C$-function. Motivated by~\cite{Casini:2012ei}, for a generic defect RG flow, it is tempting to conjecture that {\it $(\ell\partial_\ell-1) S_\mathcal{D}^{\rm p=3}$ as a defect $C$-function, is monotonically decreasing along the defect RG flows.} As previously mentioned, $S_\mathcal{D}$ stands for the defect localized EE on flat defect. Notice that the refinement we used here is the same as that used in entropic $C$-function in QFT$_3$~\cite{Liu:2012eea}. An information theoretical proof of the monotonicity of the conjectured $C$-function for $p=3$ is given in appendix~\ref{appx-ctheorem}. 

For higher dimensional defects ($p > 3$), the generalization of the above proof is not straightforward and we refer to~\cite{Casini:2017vbe, Lashkari:2017rcl} for further development of information theoretical proof of $C$-theorem in higher dimensions.

%%%%%%%%%%%%%%%%%%%%%%%%%%%%%%%%%%%%%%%%%%%%%%%%%%%%%%%%%%%%%%%%%%%%%%%%%%%%%%%%%%%%%%%%%%%%%%%
%%%%%%%%%%%%%%%%%%%%%%%%%%%%%%%%%%%%%%%%%%%%%%%%%%%%%%%%%%%%%%%%%%%%%%%%%%%%%%%%%%%%%%%%%%%%%%%
\subsection{Holographic dual}
Assuming that the bulk CFT has a holographic dual, namely an AdS gravity, one natural question is whether there exists a minimal surface such as Ryu-Takayanagi surface \cite{Ryu:2006bv} as the holographic dual of defect localized EE. We will show that such a minimal surface indeed exists provided that the defect itself has a holographic dual.

\begin{figure}[htbp]
\centering
\includegraphics[height=4.1cm]{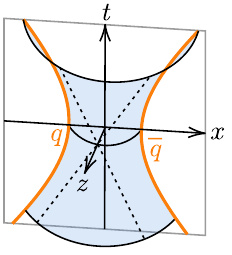}
\caption{A pair of external accelerating quarks and its holographic dual. The horizon on the worldsheet is represented by black dashed lines.}
\label{fig-quark}
\end{figure}

The first example is the $1/2$-BPS Wilson loop (WL) in $4D$ ${\cal N}=4$ SYM, which preserves a $SL(2, \mathbb{R})$ conformal subgroup and thus can be regarded as a conformal defect of the $4D$ theory. Some properties of this defect CFT have been studied in~\cite{Drukker:2006xg, Sakaguchi:2007ba, Cooke:2017qgm, Kim:2017sju, Kiryu:2018phb, Giombi:2017cqn}. Consider a circular WL in $\mathbb{R}^4$, which is conformally equivalent to a WL wrapping along a maximal circle in $\mathbb{S}^4$. As shown in figure~\ref{fig-quark}, there is a physical interpretation for the former WL in real time in terms of a pair of external accelerating quarks. According to~\cite{Rey:1998ik, Maldacena:1998im, Drukker:1999zq}, the holographic dual of the pair of quarks is a connected string in AdS$_5$ attached to the quarks at its end points. It was demonstrated in~\cite{Jensen:2013ora} that, there is an emergent wormhole in the string worldsheet, as an illustration of ER=EPR~\cite{Maldacena:2013xja}. In the large $N$ limit with fixed $\lambda \gg 1$, the strong coupling regime of WL operator is dominated by the classical string action 
\begin{equation}\label{string}
S_\text{string} = -{1 \over 2\pi \alpha'} \int \sqrt{-\det g}\ .
\end{equation}
Varying this action is equivalent to finding a minimal surface and the classical solution of the string worldsheet is AdS$_2$. From the worldsheet point of view, we can understand string/WL duality as AdS$_2$/CFT$_1$~\cite{Giombi:2017cqn}. If we simply view the on-shell string action \eqref{string} as a 2D gravity action, one can deduce an ``effective Newton constant'' (recall that the Ricci scalar of AdS$_2$ is $-2/L^2$) 
\begin{equation}\label{eq-G2}
{1\over 4 G_2}={L^2\over \alpha'}=(L/\ell_s)^2\ ,
\end{equation} 
which says nothing but the fundamental UV scale on string worldsheet is set by string length.\footnote{We do not claim a 2D gravity localized on the string worldsheet here. Instead, $4 G_2$ we find in \eqref{eq-G2} plays the role of UV fundamental area used to measure the entropy. The same thing happens in the following probe brane model.} The defect localized EE between the two quarks can be computed holographically from the horizon entropy of the wormhole
\begin{equation}\label{eq-WLentropy}
S = {A\over 4 G_2} = {L^2\over \alpha'} = \sqrt{\lambda}\ , 
\end{equation} 
where $A$ is the area of a dot fixed to be unit.
This agrees precisely with the field theory result for defect localized EE \eqref{eq:DLEE1}, which equals to the expectation value of circular 1/2-BPS WL in ${\cal N}=4$ SYM~\cite{Erickson:2000af, Drukker:2000rr, Pestun:2007rz}. Notice that the defect localized EE differs from the defect contribution to the bulk EE, worked out by Jensen and Karch~\cite{Jensen:2013ora}, as well as by Lewkowycz and Maldacena~\cite{Lewkowycz:2013laa}, which is $\sqrt{\lambda}/3$.

In~\cite{Polchinski:2011im, Beccaria:2017rbe} the authors consider a 1-parameter family of WL operators in $4D$ $\mathcal{N} = 4$ SYM
\begin{equation}\label{eq-zetaWL}
	W^{(\zeta)}(C) = \frac{1}{N} \text{Tr} \mathcal{P} \exp \oint_C d\tau \left[ i A_\mu(x) \dot{x}^\mu + \zeta \Phi_m(x)\theta^m|\dot{x}| \right]
\end{equation}
with $\theta_m^2 = 1$. The $\zeta$-dependent term in \eqref{eq-zetaWL} can be viewed as a perturbation driving a defect RG flow from the standard WL in the UV fixed point $\zeta = 0$ to the $1/2$-BPS WL in the IR fixed point $\zeta = 1$.\footnote{For further discussion on this RG flow, see~\cite{Beccaria:2018ocq, Beccaria:2021rmj, Beccaria:2022bcr}. } At weak coupling it has been verified in the Supplemental Material of~\cite{Cuomo:2021rkm} that the defect localized EE monotonically decreases along this defect RG flow. At strong coupling the logarithm expectation value of the WL is given by~\cite{Beccaria:2017rbe}\footnote{\eqref{eq-WLstrong} is slightly different from eq.(4.28) in~\cite{Beccaria:2017rbe}, where $R$ is set to 1 and absorbed into $\varkappa$.}
\begin{equation}\label{eq-WLstrong}
\begin{split}
\log \langle W^{(\varkappa)} \rangle = 
&\sqrt{\lambda} - 5\varkappa R (\log \varkappa R - 1) + 5\log \Gamma(1 + \varkappa R)\\
&- \frac{5}{2}\log(2\pi \varkappa R) - F_1(\varkappa = \infty) + \mathcal{O}(\frac{1}{\sqrt{\lambda}})\ , 
\end{split}
\end{equation}
where $R$ is the WL radius, $F_1(\varkappa = \infty)$ is $R$ independent and it is the subleading term of the logarithm expectation value of $1/2$-BPS WL calculated in~\cite{Erickson:2000af, Drukker:2000rr, Pestun:2007rz}, and $\varkappa$ is the strong coupling counterpart of $\zeta$ which is a non-trivial function of $\zeta$ and $\lambda$
\begin{equation}
\varkappa = \mathrm{f}(\zeta;\lambda)\ , \quad \mathrm{f}(0;\lambda) = 0\ , \quad \mathrm{f}(1;\lambda \gg 1) = \infty\ . 
\end{equation}
The defect localized EE can thus be calculated from \eqref{eq-WLstrong}
\begin{equation}
\begin{split}
S_D 
= &(1 - R\partial_R) \log \langle W^{(\varkappa)} \rangle\\
= &\sqrt{\lambda} + 5\varkappa R + \frac{5}{2} - \frac{5}{2}\log(2\pi \varkappa R) + 5\log \Gamma(1 + \varkappa R)\\ 
&- 5\varkappa R \frac{\Gamma'(1 + \varkappa R)}{\Gamma(1 + \varkappa R)} - F_1(\varkappa = \infty) + \mathcal{O}(\frac{1}{\sqrt{\lambda}})\ ,
\end{split}
\end{equation}
which monotonically decreases under the defect RG flow from UV ($\varkappa = 0$) to IR ($\varkappa = \infty$). 

Our second example is a holographic model of DCFT$_D$ called probe brane model, which can be viewed as the holographic dual of the defect model in figure~\ref{fig-Trans}. We work in the Euclidean coordinates after the CHM map. The probe brane has a tension $T_\text{p}$ in the Euclidean AdS space, and the Euclidean action is given by
\begin{equation}
\begin{split}
I = &- {1 \over 16\pi G_{D + 1}} \int \text{d}^{D + 1}x \sqrt{G} \left( R + {D(D - 1) \over L^2} \right)\\ 
&+ T_\text{p} \int \text{d}^{p + 1}\hat x \sqrt{\hat G}\ .
\end{split}
\end{equation}
In the probe limit, the solution of brane worldvolume is Euclidean AdS$_{p+1}$ and the defect free energy is given by the on-shell action of the brane
\begin{equation}\label{DFh}
\log \langle D[\mathbb{S}^p] \rangle = -I_{\text{brane}} = -\text{Vol}(\mathbb{H}^{p + 1}) T_\text{p} L^{p + 1}\ ,
\end{equation}
which should be equal to defect localized EE according to \eqref{SFD}. Now we give one more check for this result. The on-shell brane action can be treated as an on-shell AdS gravity action on Euclidean AdS$_{p+1}$, from which one can deduce an effective Newton constant, related to the brane tension,
\begin{equation}
T_\text{p} = {1 \over 16\pi G_{p + 1}}{2p \over L^2}\ .
\end{equation}
\begin{figure}[H]
\centering
\includegraphics[height=3.7cm]{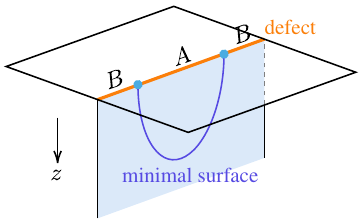}
\caption{A time slice of the $p = 2$ defect and its holographic dual. The $z$-direction is the holographic direction. }
\label{fig-probe}
\end{figure}
To compute the holographic defect localized EE, let us turn to Lorentzian coordinates. One needs to find a minimal surface within the brane worldvolume, which is a $p - 1$ dimensional minimal surface, as shown in figure~\ref{fig-probe}. Thus the holographic result can be computed by
\begin{equation}\label{DLEEh}
\begin{split}
S_D &= {A \over 4 G_{p + 1}}\\ 
&= { L^{p - 1}\text{Vol}(\mathbb{S}^{p - 2}) \over 4 G_{p + 1}}\int_{\epsilon/R}^1 \frac{(1 - y^2)^{(p - 3)/2}}{y^{p - 1}}\\ 
&={1 \over 4 G_{p + 1}}{\text{Vol}(\mathbb{H}^{p - 1}) L^{p - 1}}\ .
\end{split}
\end{equation}
Using the volume formula for hyperbolic space in general dimensions
\begin{equation}
\text{Vol}(\mathbb{H}^{p + 1}) = -{1 \over \sin(\pi p / 2)}{\pi^{p / 2 + 1} \over \Gamma(p / 2 + 1)}\ ,
\end{equation} 
we find that the defect localized EE \eqref{DLEEh} precisely agrees with the defect free energy \eqref{DFh}, a general result we proved at the fixed point of defect RG flow. Notice that for a given holographic defect, once the characteristic constant $4 G_{p + 1}$ is obtained, it can be used to compute defect localized entropy for other subregions with different shapes.
%%%%%%%%%%%%%%%%%%%%%%%%%%%%%%%%%%%%%%%%%%%%%%%%%%%%%%%%%%%%%%%%%%%%%%%%%%%%%%%%%%%%%%%%%%%%%%%
%%%%%%%%%%%%%%%%%%%%%%%%%%%%%%%%%%%%%%%%%%%%%%%%%%%%%%%%%%%%%%%%%%%%%%%%%%%%%%%%%%%%%%%%%%%%%%%
\subsection{Boundary CFT}
\begin{figure}[htbp]
\centering
\includegraphics[height=2.3cm]{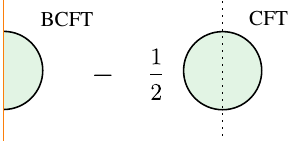}
\caption{Interpretation of boundary localized entropy.}
\label{fig-bdyE}
\end{figure}

The boundary can be viewed as a special codimension one defect. We define boundary localized entropy by slightly modifying \eqref{SD} 
\begin{equation}\label{eq-bdySD}
    S_D^\text{bdy} = \log \frac{Z^\text{BCFT}}{\sqrt{Z^\text{CFT}}} + \int_{\mathbb{S}^p} \llangle t^\tau{}_\tau\rrangle\ , 
\end{equation}
where the square root comes from the normalization of BCFT partition function on hemisphere.\footnote{Several results on boundary contributions to the Weyl anomaly have been given for $2D$~\cite{Polchinski:1998rq}, $3D$~\cite{Jensen:2018rxu}, $4D$~\cite{Herzog:2015ioa, Herzog:2017kkj} and $5D$~\cite{FarajiAstaneh:2021foi, Chalabi:2021jud}.} Notice that in $D = 2$ this definition is consistent with the boundary entropy defined in~\cite{Friedan:2003yc}. One can interpret the boundary localized entropy as the difference between the entanglement entropy with and without the boundary,\footnote{\eqref{bCFTSD} can also be expressed as 
\begin{small}
\begin{equation*}
\begin{split}
S_D^\text{bdy} = 
&(1 - n\partial_n) \log {Z^\text{BCFT}_n \over \sqrt{Z_n^\text{CFT}}}\big|_{n \to 1}\\
=&\log {Z^\text{BCFT} \over \sqrt{Z^\text{CFT}}} + \int_{\mathbb{HS}^D} \llangle T^\tau{}_\tau + \delta(\phi) t^\tau{}_\tau \rrangle - \frac{1}{2} \int_{\mathbb{S}^D} \llangle T^\tau{}_\tau \rrangle\ . 
\end{split}
\end{equation*}
\end{small}
Canceling out the bulk stress tensor $T^\tau{}_\tau$ gives \eqref{eq-bdySD}. 
} i.e.
\begin{equation}\label{bCFTSD}
    S_D^\text{bdy} = S_\text{EE}^\text{BCFT} - \frac{1}{2} S_\text{EE}^\text{CFT}
\end{equation}
as illustrated in figure~\ref{fig-bdyE}. 

In $D = 2$, the boundary localized entropy as a $g$-function is monotonically decreasing, first conjectured by Affleck and Ludwig~\cite{Affleck:1991tk, Ludwig:1994nf} and then proved by Friedan and Konechny~\cite{Friedan:2003yc}. An alternative proof of the $g$-theorem was given in~\cite{Casini:2016fgb} using the monotonicity of relative entropy. For holographic discussions we refer to~\cite{Yamaguchi:2002pa, Fujita:2011fp, Erdmenger:2013dpa, Erdmenger:2015spo, Erdmenger:2020hug}.

Now we consider a holographic model for BCFT$_D$ as an AdS gravity plus a bounding brane with constant tension proposed by Takayanagi~\cite{Takayanagi:2011zk} and calculate the boundary localized entropy holographically. The action of this holographic system is given by
\begin{equation}
I = - {1 \over 16\pi G_{D + 1}} \int \sqrt{g} (R - 2\Lambda) - {1 \over 8\pi G_{D+1}} \int \sqrt{h}(K - T)\ ,
\end{equation}  
with $T$ the brane tension. The solution of the bulk metric can be solved,
\begin{equation}
\text{d}s^2 = \text{d}\rho^2 + L^2 \cosh^2{\rho \over L} \left( \frac{-\text{d}t^2 + \text{d}y^2 + \text{d} \vec x^2}{y^2}\right)\ ,
\end{equation}
with the brane located at $\rho = \rho_*$ and $\rho_*$ is determined by the brane tension through Neumann boundary conditions. We choose a subregion surrounded by a $D - 3$ dimensional sphere with radius $R$ within the boundary defect.  
Following \eqref{bCFTSD}, the holographic dual for boundary localized entropy will be the minimal surface within the wedge between $\rho=0$ and $\rho=\rho_*$. The computation of the area of this minimal surface is given by
\begin{equation}
S = {L^{D - 2} \text{Vol} (\mathbb{S}^{D - 3}) \over 4G_{D + 1}} \int_{\epsilon/R}^1 \frac{(1 - y^2)^{D/2 - 2}}{y^{D - 2}} \int_0^{\rho_*} \cosh^{D - 2} {\rho \over L} \ , 
\end{equation}  
which agrees with the boundary entropy in~\cite{Kobayashi:2018lil}.

%%%%%%%%%%%%%%%%%%%%%%%%%%%%%%%%%%%%%%%%%%%%%%%%%%%%%%%%%%%%%%%%%%%%%%%%%%%%%%%%%%%%%%%%%%%%%%%
%%%%%%%%%%%%%%%%%%%%%%%%%%%%%%%%%%%%%%%%%%%%%%%%%%%%%%%%%%%%%%%%%%%%%%%%%%%%%%%%%%%%%%%%%%%%%%%
\subsection{Conclusion and Discussion}
In this paper we propose defect localized entropy as a measure of degrees of freedom on defect. The defect localized entropy is defined by performing replica trick only on the defect and keeping the bulk intact. It counts the entanglement between the degrees of freedom localized on the defect. It is demonstrated that at the defect fixed point, defect localized entropy is equal to minus defect free energy. We construct defect $C$-functions based on defect localized entropy and show that they monotonically decrease for surface defects and volume defects. We also study the holographic dual of defect localized entropy for several examples. 

A few future questions are listed in order: First, it is interesting to explore more examples of defect RG flows as well as their holographic duals and to verify the monotonically decreasing behavior of defect localized entropy (or its refinement). Second, generalize the field theoretical argument of the monotonicity for line defects~\cite{Cuomo:2021rkm}. Last but not least, explore other physical applications of this newly defined defect localized entropy in particle physics as well as in condensed matter physics.
%%%%%%%%%%%%%%%%%%%%%%%%%%%%%%%%%%%%%%%%%%%%%%%%%%%%%%%%%%%%%%%%%%%%%%%%%%%%%%%%%%%%%%%%%%%%%%%
%%%%%%%%%%%%%%%%%%%%%%%%%%%%%%%%%%%%%%%%%%%%%%%%%%%%%%%%%%%%%%%%%%%%%%%%%%%%%%%%%%%%%%%%%%%%%%%
\subsection*{Acknowledgments}
We are grateful for useful discussions with Shimon Yankielowicz and our group members in Fudan University. We thank Xinan Zhou and Junbao Wu for reading this manuscript. We also thank Horacio Casini, Nadav Drukker, Chris Herzog and Zohar Komargodski for useful correspondence. This work is supported by NSFC grant 11905033. Part of this work is presented in SUIAS workshop on Supersymmetry and Gravitation as well as in the 3rd national workshop on QFT and String Theory in China.
%%%%%%%%%%%%%%%%%%%%%%%%%%%%%%%%%%%%%%%%%%%%%%%%%%%%%%%%%%%%%%%%%%%%%%%%%%%%%%%%%%%%%%%%%%%%%%%
%%%%%%%%%%%%%%%%%%%%%%%%%%%%%%%%%%%%%%%%%%%%%%%%%%%%%%%%%%%%%%%%%%%%%%%%%%%%%%%%%%%%%%%%%%%%%%%
\appendix
\section{\boldmath Path integral construction of the reduced density matrix \unboldmath}\label{appx-PI}
First we write down the DCFT action \eqref{eq:action} more precisely, 
\begin{align}
I_{\text{CFT}} &= \int \text{d}^D x \sqrt{g} {\cal L}_{\text{CFT}}[\phi(x)]\ , \\
I_{\text{defect}} &= \int \text{d}^p \hat x \sqrt{\gamma} {\cal L}_{\text{defect}}[\phi(\hat{x}), \psi(\hat{x})]\ . 
\end{align}
Here we used $\phi$ to represent the bulk degrees of freedom and $\psi$ the defect degrees of freedom. We construct the following reduced density matrix by cutting along the defect subregion $A$ (the entire defect at $t=0$ is $A \cup B$) and performing path integral over the full spacetime\footnote{The unnormalized reduced density matrix should be
\begin{equation}\label{eq-unnormal}
\begin{split}
[\hat{\rho}'_A]_{ab} = &\frac{1}{Z^\text{CFT}} \int_{\cal{M}} \mathcal{D}\phi \int_{\Sigma} \mathcal{D}\psi\ e^{- I_{\text{CFT}} - I_{\text{defect}}}\\
&\times \prod_{\hat{x} \in A} \delta (\psi(0^+, \hat{x}) - \psi_b(\hat{x})) \delta (\psi(0^-, \hat{x}) - \psi_a(\hat{x}))\ , 
\end{split}
\end{equation}
for the reason that when the bulk and defect degrees of freedom are decoupled, this unnormalized density matrix can be reduced to the unnormalized density matrix with only defect degrees of freedom. }
\begin{equation}\label{eq:rhohat-A}
\begin{split}
[\hat{\rho}_A]_{ab} = &\frac{1}{Z^\text{DCFT}} \int_{\cal{M}} \mathcal{D}\phi \int_{\Sigma} \mathcal{D}\psi\ e^{- I_{\text{CFT}} - I_{\text{defect}}}\\ 
&\times \prod_{\hat{x} \in A} \delta (\psi(0^+, \hat{x}) - \psi_b(\hat{x})) \delta (\psi(0^-, \hat{x}) - \psi_a(\hat{x}))\ , 
\end{split}
\end{equation}
%\begin{equation}\label{eq:rhohat-A}
%\begin{split}
%[\hat{\rho}_A]_{ab} = &\frac{1}{Z^\text{DCFT}} \int [\mathcal{D}\phi] [\mathcal{D}\psi] e^{- I_{\text{CFT}} - I_{\text{defect}}}\\
%&\cdot \prod_{\hat{x} \in A} \delta (\psi(0^+, \hat{x}) - \psi_b(\hat{x})) \delta (\psi(0^-, \hat{x}) - \psi_a(\hat{x}))\ , 
%\end{split}
%\end{equation}
where the indices $a$ and $b$ denote different configurations of the field $\psi$. The trace of the $n$-th power of this reduced density matrix gives $n$ copies of the bulk and glues the defect into a $n$-sheeted surface, i.e.
\begin{equation}\label{eq:tr-rho-n}
\begin{split}
\text{Tr} \hat{\rho}_A^n = &\frac{1}{(Z^\text{DCFT})^n} \int_{\mathcal{M}_{1}} \mathcal{D}\phi_1 \cdots \int_{\mathcal{M}_n} \mathcal{D}\phi_n \int_{\Sigma^{(n)}} \mathcal{D}\psi\\ 
&\times e^{- \sum_{i=1}^{n} I_{\text{CFT}}[\phi_i] - I^{(n)}_{\text{defect}}[\phi_1,\dots,\phi_n,\psi]}\ , 
\end{split}
\end{equation}
%\begin{equation}\label{eq:tr-rho-n}
%\text{Tr} \hat{\rho}_A^n = \frac{1}{(Z^\text{DCFT})^n} \int [\mathcal{D}\phi]^n [\mathcal{D}^{(n)}\psi] e^{- n I_{\text{CFT}} - I^{(n)}_{\text{defect}}}\ , 
%\end{equation}
where $(n)$ denotes $n$-sheeted surface. Now we consider the case that the glued defect can receive the same bulk averaging from a single bulk copy. This happens when the bulk number of degrees of freedom is much greater than that of the defect. Therefore, \eqref{eq:tr-rho-n} can also be understood as the partition function with the $n$-fold cover of the original defect in the first bulk copy, thus the path integral on the remaining $(n-1)$ bulk copies can be factorized, which gives 
\begin{equation}
\begin{split}
\text{Tr} \hat{\rho}_A^n 
&= \frac{(Z^\text{CFT})^{n-1}}{(Z^\text{DCFT})^n} \int_{\cal{M}} \mathcal{D}\phi \int_{\Sigma^{(n)}} \mathcal{D}\psi\ e^{- I_{\text{CFT}} - I^{(n)}_{\text{defect}}}\\ 
&= \left( \frac{Z^\text{CFT}}{Z^\text{DCFT}} \right)^n \frac{\int_{\cal{M}} \mathcal{D}\phi \int_{\Sigma^{(n)}}\mathcal{D}\psi\ e^{- I_{\text{CFT}} - I^{(n)}_{\text{defect}}}}{\int_{\cal{M}} \mathcal{D}\phi\ e^{- I_{\text{CFT}}}}\\ 
&= \frac{\left< D_n \right>}{\left< D \right>^n}\ . 
\end{split}
\end{equation}
%\begin{equation}
%\begin{split}
%\text{Tr} \hat{\rho}_A^n 
%&= \frac{(Z^\text{CFT})^{n-1}}{(Z^\text{DCFT})^n} \int [\mathcal{D}\phi] [\mathcal{D}^{(n)}\psi] e^{- I_{\text{CFT}} - I^{(n)}_{\text{defect}}}\\ 
%&= \left( \frac{Z^\text{CFT}}{Z^\text{DCFT}} \right)^n \frac{\int [\mathcal{D}\phi] [\mathcal{D}^{(n)}\psi] e^{- I_{\text{CFT}} - I^{(n)}_{\text{defect}}}}{\int [\mathcal{D}\phi] e^{- I_{\text{CFT}}}}\\ 
%&= \frac{\left< D_n \right>}{\left< D \right>^n}\ . 
%\end{split}
%\end{equation}
Therefore we rederive \eqref{SD0} from the reduced density matrix \eqref{eq:rhohat-A}, 
\begin{equation}
S_{D} = \lim_{n \rightarrow 1} \frac{1}{1 - n} \log\frac{\left< D_n \right>}{\left< D \right>^n} = \lim_{n \rightarrow 1} \frac{1}{1 - n} \log \text{Tr} \hat{\rho}_A^n\ .
\end{equation}
%%%%%%%%%%%%%%%%%%%%%%%%%%%%%%%%%%%%%%%%%%%%%%%%%%%%%%%%%%%%%%%%%%%%%%%%%%%%%%%%%%%%%%%%%%%%%%%
%%%%%%%%%%%%%%%%%%%%%%%%%%%%%%%%%%%%%%%%%%%%%%%%%%%%%%%%%%%%%%%%%%%%%%%%%%%%%%%%%%%%%%%%%%%%%%%
\section{\boldmath An alternative derivation of \eqref{SD} \unboldmath}\label{appx-CHM}
In this appendix we give an alternative derivation of \eqref{SD} based on the density matrix \eqref{eq:rhohat-A} and the original CHM map method in~\cite{Casini:2011kv}. The reduced density  matrix \eqref{eq:rhohat-A} is an averaged density matrix, which is still hermitian and positive semi-definite (also note that it has been normalized). After the CHM map \eqref{coordT}, \eqref{eq:rhohat-A} is transformed into a thermal density matrix, which can be expressed as 
\begin{equation}
\hat{\rho} = \frac{e^{-\beta H_\tau}}{Z^\text{DCFT}/Z^\text{CFT}} = \frac{1}{\langle D \rangle} e^{-\beta H_\tau}\ , 
\end{equation}
with $H_\tau$ the infinitesimal generator of the defect $\tau$ translation and the normalization factor determined from the unnormalized density matrix \eqref{eq-unnormal}. Following eq.(4.6) in~\cite{Casini:2011kv}, we have
\begin{equation}
\begin{split}
S_D 
&= -\text{Tr} \left( \hat{\rho} \log \hat{\rho} \right)\\
&= \log \langle D \rangle + \beta \text{Tr} \left(\hat{\rho} H_\tau \right)\ , 
\end{split}
\end{equation}
where the second term is the expectation value of the operator generating the defect $\tau$ translation, which is just the Killing energy (times $\beta = 2 \pi \ell$) localized on the defect. Therefore, we obtain \eqref{SD} from the density matrix \eqref{eq:rhohat-A} by CHM map.
%%%%%%%%%%%%%%%%%%%%%%%%%%%%%%%%%%%%%%%%%%%%%%%%%%%%%%%%%%%%%%%%%%%%%%%%%%%%%%%%%%%%%%%%%%%%%%%
%%%%%%%%%%%%%%%%%%%%%%%%%%%%%%%%%%%%%%%%%%%%%%%%%%%%%%%%%%%%%%%%%%%%%%%%%%%%%%%%%%%%%%%%%%%%%%%
\section{\boldmath monotonicity of $p = 3$ defect $C$ function\unboldmath}\label{appx-ctheorem}
\begin{figure}[htbp]
\centering
\includegraphics[height=3.4cm]{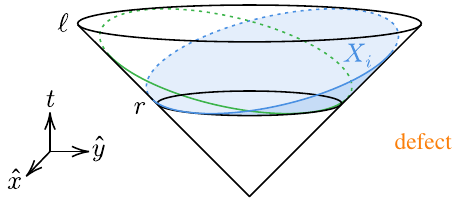}
\caption{A proof of the $p = 3$ defect $C$-theorem \textbf{(working on defect)}. This solid cone is the causal past of the $\ell$-disk and is placed on the \textbf{defect}. A plane parallel to the $\hat{x}O\hat{y}$-plane and the causal past intersect a disk of radius $r$. We have a series of $N$ boosted circles of radius $\sqrt{r\ell}$ (here we draw two of them) tangent to the $r$-circle and the $\ell$-circle. $X_i$ denotes the blue shaded area whose boundary is the $i$-th boosted circle. }
\label{fig-cone}
\end{figure}

Now we provide an information theoretical proof of the monotonicity of $p = 3$ defect $C$ function following~\cite{Casini:2012ei}. Similar to the proof for $p=2$, we trigger the defect RG flow by tuning the system size $\ell$. The defect $C$-theorem is then stated as: $c(\ell)$ is a monotonically decreasing function under RG flows. As illustrated in figure~\ref{fig-cone}, consider a disk of radius $\ell$ located on a time slice of defect with its causal past a solid cone. A plane parallel to the $\hat{x}O\hat{y}$-plane intersects this solid cone with a disk of radius $r$. Then we intersect the light cone (the boundary of the solid cone) with a series of $N$ planes to get a series of $N$ boosted circles of radius $\sqrt{r\ell}$. These boosted circles are tangent to the two circles of radius $r$ and $\ell$, and the angle between the the projections of adjacent boosted circles on the $\hat{x}O\hat{y}$-plane is $2\pi/N$. We use $X_i$ ($i = 1, \dots, N$) to denote the blue shaded area in figure~\ref{fig-cone}, which is the union of the $r$-disk and a part of the light cone bounded by the $i$-th boosted circle and the $r$-circle. 

For these $X_i$ we apply the following inequality which is obtained by repeatedly using the strong subadditivity of defect localized EE
%\begin{small}
\begin{equation}\label{eq-appxSSA}
\begin{split}
\sum_i S_\mathcal{D}(X_i) \geq\ &S_\mathcal{D}(\cup_i X_i) + S_\mathcal{D}(\cup_{\{ij\}} X_{ij})\\ 
&+ S_\mathcal{D}(\cup_{\{ijk\}} X_{ijk}) + \cdots + S_\mathcal{D}(\cap_i X_i)\ ,
\end{split}
\end{equation}
%\end{small}
where $X_{ij\cdots k}$ denotes $X_i \cap X_j \cap \cdots \cap X_k$ with no repeated indices $i,j,\dots,k$. As $N$ goes to infinity, the boundary of area $\cup_{\{ij\dots k\}} X_{ij\cdots k}$ approaches a circle of radius $r_n$, where $n$ is the number of indices and $r_n$ can be obtained by geometric calculation
\begin{equation}\label{eq-OPn}
r_n = \frac{2r\ell}{r + \ell + (r - \ell)\cos\frac{n\pi}{N}}\ . 
\end{equation}
Due to the unitarity and Lorentz invariance, when $N$ goes to infinity, $S_D(\cup_{\{ij\cdots k\}} X_{ij\cdots k})$ can be directly replaced by the defect localized EE of a disk of radius $r_n$. Substituting \eqref{eq-OPn} into \eqref{eq-appxSSA} and taking $N \to \infty$ gives 
\begin{equation}\label{eq-sum}
\begin{split}
S_\mathcal{D}(\sqrt{r\ell}) &\geq \lim_{N \to \infty} \frac{1}{N} \sum_{n = 1}^{N} S_\mathcal{D} \left( \frac{2r\ell}{r + \ell + (r - \ell)\cos\frac{n\pi}{N}} \right)\\
&= \frac{1}{\pi} \int_0^\pi dz\ S_\mathcal{D}\left( \frac{2r\ell}{r + \ell + (r - \ell)\cos z} \right)\ . 
\end{split}
\end{equation}
Performing Taylor expansion of \eqref{eq-sum} at $\ell$ with respect to $\epsilon = r - \ell$ to the second order, we obtain 
\begin{equation}
\begin{split}
&S_\mathcal{D}(\ell) + \left( \frac{\epsilon}{2} - \frac{\epsilon^2}{8\ell} \right) S'_\mathcal{D}(\ell) + \frac{\epsilon^2}{8} S''_\mathcal{D}(\ell)\\ 
\geq\ &S_\mathcal{D}(\ell) + \left( \frac{\epsilon}{2} - \frac{\epsilon^2}{8\ell} \right) S'_\mathcal{D}(\ell) + \frac{3\epsilon^2}{16} S''_\mathcal{D}(\ell)\ , 
\end{split}
\end{equation}
thus we complete the proof
\begin{equation}
\left( (\ell \partial_\ell - 1) S_\mathcal{D} \right)' = \ell S''_\mathcal{D}(\ell) \leq 0\ . 
\end{equation}
%%%%%%%%%%%%%%%%%%%%%%%%%%%%%%%%%%%%%%%%%%%%%%%%%%%%%%%%%%%%%%%%%%%%%%%%%%%%%%%%%%%%%%%%%%%%%%%
%%%%%%%%%%%%%%%%%%%%%%%%%%%%%%%%%%%%%%%%%%%%%%%%%%%%%%%%%%%%%%%%%%%%%%%%%%%%%%%%%%%%%%%%%%%%%%%
\bibliography{DLEE}

\end{document}